\documentclass[aps,prb,reprint,showpacs,amsmath,amssymb]{revtex4-1}
\usepackage{units}
\usepackage{graphicx}

\begin{document}

\title{Anomalous Surface Segregation Profiles in Ferritic FeCr Stainless Steel}

\author{Maximilien Levesque}
\email{maximilien.levesque@ens.fr}
\affiliation{\'Ecole Normale Sup\'erieure, D\'epartement de Chimie, UMR 8640 CNRS-ENS-UPMC, 24 rue Lhomond, 75005 Paris, France}
\affiliation{Universit\'e Pierre et Marie Curie, CNRS UMR 7195 PECSA, 75005, Paris, France}

\begin{abstract}
The iron-chromium alloy and its derivatives are widely used for their
remarkable resistance to corrosion, which only occurs in a narrow concentration
range around 9 to 13 atomic percent chromium. Although known to be
due to chromium enrichment of a few atoms thick layer at the surfaces, the
understanding of its complex atomistic origin has been a remaining
challenge. We report an investigation
of the thermodynamics of such surfaces at the atomic scale by means
of Monte Carlo simulations. We use a Hamiltonian which provides a parameterization of previous \emph{ab initio} results
and successfully describes the alloy's unusual thermodynamics.
We report a strong enrichment
in Cr of the surfaces for low bulk concentrations, with a narrow optimum
around 12 atomic percent chromium, beyond which the surface composition decreases drastically.
This behavior is explained by a synergy between
(i)~the complex phase separation in the bulk alloy,
(ii)~local phase transitions that tune the layers closest to the surface to an iron-rich state and inhibit the
bulk phase separation in this region, and
(iii)~its compensation
by a strong and non-linear enrichment in Cr of the next few layers.
Implications with respect to the design of prospective nanomaterials are briefly discussed.
\end{abstract}


\maketitle

The iron-chromium alloy and its derivatives are inexpensive, have
satisfactory mechanical properties and above all exhibit a remarkable resistance
to corrosion: it is the most widely used class of alloy in the world.
Its outstanding corrosion resistance is known for a century~\cite{Monnartz1911} to only occur in a narrow range of
concentrations, around 10 atomic percent of chromium
(at.~\%~Cr)~\cite{Asami78CrXRay}. Their excellent properties make
them candidate materials for future fusion nuclear reactors~\cite{odette_annual_review_2008,Dudarev2009EUprogram},
one of the reasons that induced a considerable amount of work on the
various aspects of the Fe--Cr alloy both experimentally~\cite{Bonny2008rev,Xiong2010rev}
and theoretically~\cite{BonnyStateArtModels09}.

Corrosion resistance of stainless steels is due to the passivation
of the material by an inert, chromium rich layer at the interface
between the alloy and the environment, \emph{i.e.} at the surfaces. Passivation
is a phenomena inherent to how much Cr is located at the surfaces, which
is a non-linear function of the bulk concentration~\cite{book_corrosion_1996}. In austenitic Fe--Cr, which only exists at high temperatures above $\approx 800$~°C and for less than $\approx 10$~at.~\%~Cr, the more chromium in the bulk, the more chromium in the surface and thus the more stainless the alloy. In ferritic Fe--Cr alloys, the picture is more complex. Without additive elements, the Cr content at which the alloy is passivated is narrow, from 9 to 13~at.~\%~Cr, beyond which occurs an increase in the corrosion rate and a strong decrease in mechanical properties.

This important property of stainless steels has been extensively studied,
but its complex origin at the atomic scale has remained a missing understanding, subject to controversial findings:
How chromium causes passivation, \emph{i.e.} how it interacts and reacts with chemical elements coming from the environment like dioxygen or hydrogen \cite{lev_gupta_fecrh}, is out of the scope of this study. The reader is refered to Greeley et al.~\cite{greeley_electronic_2002} for a review of surface chemistry of metal surfaces at the atomic and electronic scale. Surface reaction requires nevertheless that Cr is present in large enough quantity on the surface to form a few atoms thick protective layer, \emph{e.g.} of chromium(III) oxide Cr$_2$O$_3$.
How chromium atoms enrich the surfaces remains unclear.
Venus and Heinrich~\cite{venus_interfacial_1996} shew by angle-resolved Auger electron spectroscopy
that Cr atoms deposited on a whisker of Fe~(100) migrates from the surface to the first few layers, in contradiction with the expected tendency.
This surface-alloying has been clearly identified to be linked to anomalies in the magnetic properties of the Cr/Fe system,
specifically the change in surface magnetization at low Cr coverage and the strong interactions between surface Cr atoms~\cite{unguris_magnetism_1992,davies_prl96}.
Ropo\emph{~et~al.} showed by First Principles that a pure Fe--Cr surface
behaves like stainless steels with respect to Cr enrichment~\cite{Ropo07}.
They also put in
evidence a competition between the relative stabilities of the surfaces and the complex
thermodynamics of the bulk alloy.
Later, \emph{ab initio} calculations revealed that unexpected interactions between subsurface Cr atoms and
surface Fe atoms~\cite{LevCrseg} are at the origin of an anomalous~\cite{Ponomareva07,Kiejna08}
segregation behavior of Cr in Fe in the dilute regime.

At temperatures of industrial and technological interest, \emph{i.e.} between
300 and 600~K, the body-centered cubic (bcc) solid-solution of Fe--Cr
shows a miscibility gap from $9-13$ to $94-99$~at.~\%~Cr~\cite{Xiong2010rev}.
Inside, a phase-separation occurs into an iron-rich bcc solid solution,
$\alpha$, and chromium-rich bcc precipitates, $\alpha^{\prime}$, as
one would expect for a binary alloy that seemed to have a segregation tendency, \emph{i.e.} that mix
solely for entropic reasons. However, both theoretical and experimental
studies subverted this simple picture, showing favorable dissolution
energy of chromium in iron up to an anomalously high $\approx7$~at.~\%~Cr~\cite{hennion_theorySRO_1983,mirebeau_PRLfirst_1984,olsson_abFeCr1st_2003}
due to a competition between repulsive Cr-Cr interactions and attractive
Fe-Cr interactions~\cite{Ackland_MagnetImmisc,Olsson_ElectronicOrigin,Klaver_MagneticOrigin,Froideval_ExpOrigin}.
The increased chromium content leads to more frustrated magnetic interactions
in between Cr atoms that make the dissolution exothermic at low concentrations, then endothermic. Several
theoretical models~\cite{CaroCDM,Olsson2BM,BonnyStateArtModels09,NguyenDudi_StonerPot09,BonnyPotential2011,LevFeCr}
have successfully reproduced the sign change of the mixing energy.

Here, the iron-chromium ferritic stainless steel is modeled
by the Hamiltonian proposed recently in Ref.~\cite{LevFeCr}. It is specifically
designed to reproduce both (i) the whole experimental $\alpha$--$\alpha^\prime$ phase diagram
at all temperatures and compositions, and (ii) the change of sign of the mixing energies.
It is also compatible with large-scaled simulations, because of its conceptual simplicity and its rigid bcc lattice nature:
 The internal energy $\Delta H_{\text{mix}}=-\Omega c_{b}\left(1-c_{b}\right)$ is a function of the bulk concentration in chromium, $c_{b}$, 
and of the order energy $\Omega$ described in terms of local-concentration and temperature dependent pair interactions:
\begin{equation}
\Omega=\sum_{i}\frac{z^{\left(i\right)}}{2}\left(\epsilon_{AA}^{\left(i\right)}+\epsilon_{BB}^{\left(i\right)}-2\epsilon_{AB}^{\left(i\right)}\right),\label{eq:order_energy_pairs}
\end{equation}
where $z^{\left(i\right)}$ is the coordination number of shell $i$ and
$\epsilon_{jj^{\prime}}^{\left(i\right)}$ is the pair interaction
between atoms of type $j$ and $j^{\prime}$ on $i$th neighbor sites.
Influenced by the strategy of Caro~et~al.~\cite{CaroCDM}, the
order energy is advantageously expressed as a simple concentration
and temperature dependent Redlich-Kister~\cite{Redlich} expansion:
\begin{equation}
\Omega\left(x,T\right)=\left(x-\alpha\right)\left(\beta x^{2}+\gamma x+\delta\right)\left(1-\frac{T}{\theta}\right),\label{eq:order_energy_poly}
\end{equation}
where $x$ is the local concentration and $T$ the temperature.
Discrete mixing energies have been calculated \emph{ab~initio} in the whole range of concentrations and interpolated by Eq.~\ref{eq:order_energy_poly}, whose coefficients $\alpha,\,\beta,\,\gamma,\,\delta$ are given in Table~\ref{tab:Parameters_of_potential}. Coefficient $\theta$, also given in Table~\ref{tab:Parameters_of_potential}, is the critical temperature of the miscibility gap.

Homo-atomic pair interaction energies in Eq.~\ref{eq:order_energy_pairs}
are given by the experimental cohesive energy of the pure elements, given in Tab.~\ref{tab:Parameters_of_potential}, according to
$E_{\text{coh}}\left(j\right)=-\sum_{i}z^{\left(i\right)}\epsilon_{jj}^{\left(i\right)}$.
The expressions of the hetero-atomic interactions $\epsilon_{AB}^{(i)}\left(x,T\right)$
are then easily deduced from Eq.~\ref{eq:order_energy_pairs} and \ref{eq:order_energy_poly}.
They are consequently simple parametric functions of the temperature $T$ and local concentration $x$.
This last quantity, $x$, around a pair including
an atom on site $i$ and an atom on another site $j$ is naturally
defined as
\begin{eqnarray}
x & = & \frac{\sum_{n=0}^{r}\sum_{k=1}^{z^{\left(n\right)}}p_{ik}^{\left(n\right)}+\sum_{n=0}^{r}\sum_{k=1}^{z^{\left(n\right)}}p_{jk}^{\left(n\right)}}{2\sum_{n=0}^{r}z^{\left(n\right)}},\label{eq:definition_of_local_concentration}
\end{eqnarray}
where $p_{ik}^{\left(n\right)}=1$ when the $k$th site of the $n$th
coordination shell of site $i$ is a Cr atom, and 0 if it is a Fe
atom or an empty site, \emph{i.e.} a site outside the surface. The
interaction range is restricted to second nearest neighbors with $\epsilon_{ij}^{\left(2\right)}=\epsilon_{ij}^{\left(1\right)}/2$,
which has been found optimal. It is worth emphasizing that the resulting
bulk phase diagram is in very good agreement with the most recent
experimental reviews~\cite{Bonny2008rev,Xiong2010rev}: While this model does not capture the extraordinarily complex electronic structure of the bcc Fe--Cr alloys, it captures both the local nature of the interactions and the associated energetics, without empirical parameters.

\begin{table}
\begin{centering}
\begin{tabular}{|c|c|c|c|c||c|c|}
\hline 
$\alpha$ & $\beta$~(eV) & $\gamma$~(eV) & $\delta$~(eV) & $\theta$~(K) & $E_\text{coh.}$(Fe) & $E_\text{coh.}$(Cr) \\
\hline 
\hline 
0.070 & $-2.288$ & 4.439 & $-2.480$ & 1400 & 4.28 & 4.10 \\
\hline 
\end{tabular}
\par\end{centering}

\caption{\label{tab:Parameters_of_potential}Parameters of the local-concentration
and temperature dependent pair potential from Ref.~\cite{LevFeCr}. Experimental cohesive energies from Ref.~\cite{kittel} are given in eV per atom.}
\end{table}

\begin{figure}
\begin{centering}
\includegraphics[width=0.70\columnwidth]{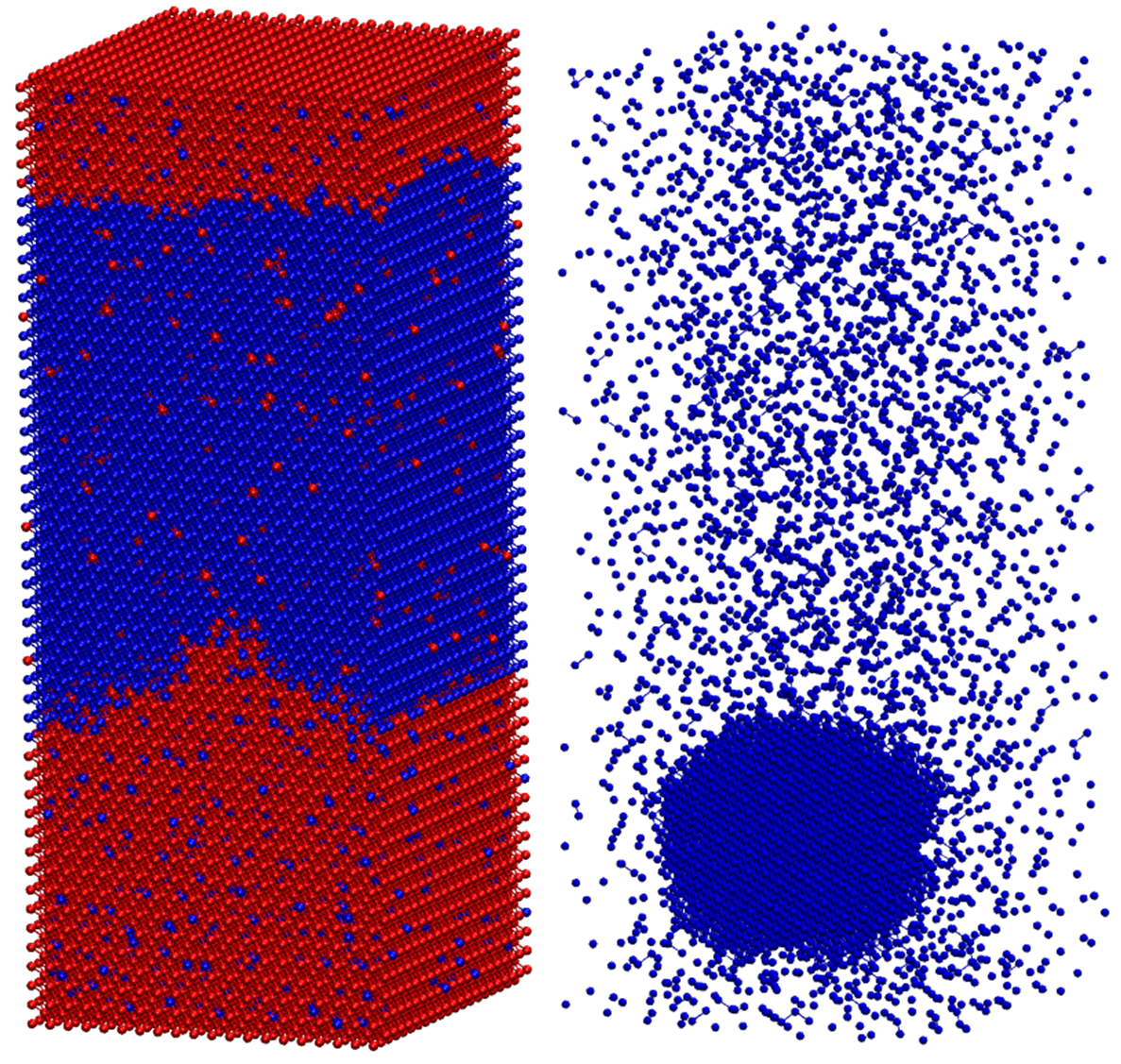}
\par\end{centering}

\caption{\label{fig:supercell}Snapshots of the simulation cell containing
the (100) FeCr surface at 300~K for bulk concentrations $c_{b}=0.5$
(left) and $c_{b}=0.15$ (right). The top surface layer is at the top of the figure. Fe atoms are shown in red and Cr
atoms in blue. For $c_{b}=0.15$, only Cr atoms are shown. In both
cases, the bulk phase separation $\alpha-\alpha^{\prime}$ occurs.}
\end{figure}

This letter focuses on the most stable surface of bcc iron, which has the orientation (100)~\cite{AldenSurfEnergies3d94}.
It is modeled by a stack of
100 layers of 400 atoms each, in periodic boundary conditions. Interactions
between periodic images in direction $\left\langle 100\right\rangle $
are prevented by a slab of vacuum. Special attention has been given to the choice of the size of the system in order (i) not to artificially hide the surface effects by a too large volume/surface ratio, and (ii) not to restrain the formation of precipitates by too small systems, ie to give the system the ability to precipitate and have the precipitates to interact with the surfaces. It induces that the bulk solubility limit near the surfaces can be slightly different from that of a pure bulk system.
A perspective view of the supercell is shown in Fig.~\ref{fig:supercell}.a.

As stated above, the model does not capture the effect of the surface on the electronic structure
of atoms in its vicinity, such as expected stronger bonds. It captures the effect of the reduced coordination in terms of energetics: Eq.~\ref{eq:definition_of_local_concentration} implies that the more reduced the coordination, the more the local energetics are dependent on the remaining bonds.
The various surface orientations only differ in the number of surface-induced dangling bonds, and thus in the strength of the surface effects described below.
Conclusions are thus transferable to other orientations.
Importantly, the effectiveness of our Hamiltonian allows to deal
with a number of atoms that make it possible to finely tune the bulk and layer
concentrations. Here, one atom accounts for the bulk
concentration by less than $10^{-4}$~at.~\% and the layer concentration by $5\cdot10^{-3}$~at.~\%.
This point is crucial as recent \emph{ab~initio} calculations
have been limited to few layers and few atoms per layer, imposing
large bulk and even larger planar concentrations. Our Hamiltonian
is sampled by Monte Carlo simulations using the Metropolis algorithm
in the canonical and pseudo grand-canonical ensembles~\cite{ducastelle_order_1991}.
The equilibrium state is considered reached after $10^{4}$ accepted
permutations per site.

We define the planar concentration $c_{p}=\sum_{i=1}^{N_{p}}q_{i}/N_{p}$
as the chromium content of each layer $p$ parallel to the surface,
with $N_{P}$ the number of atoms per layer (400 here) and $q_{i}=1$
if site $i\subset p$ contains a Cr atom, $q_{i}=0$ otherwise. $p$ ranges
from 0 for the top surface layer to $99$ for the bottom surface layer.

\begin{figure}
\begin{centering}
\includegraphics[width=1\columnwidth]{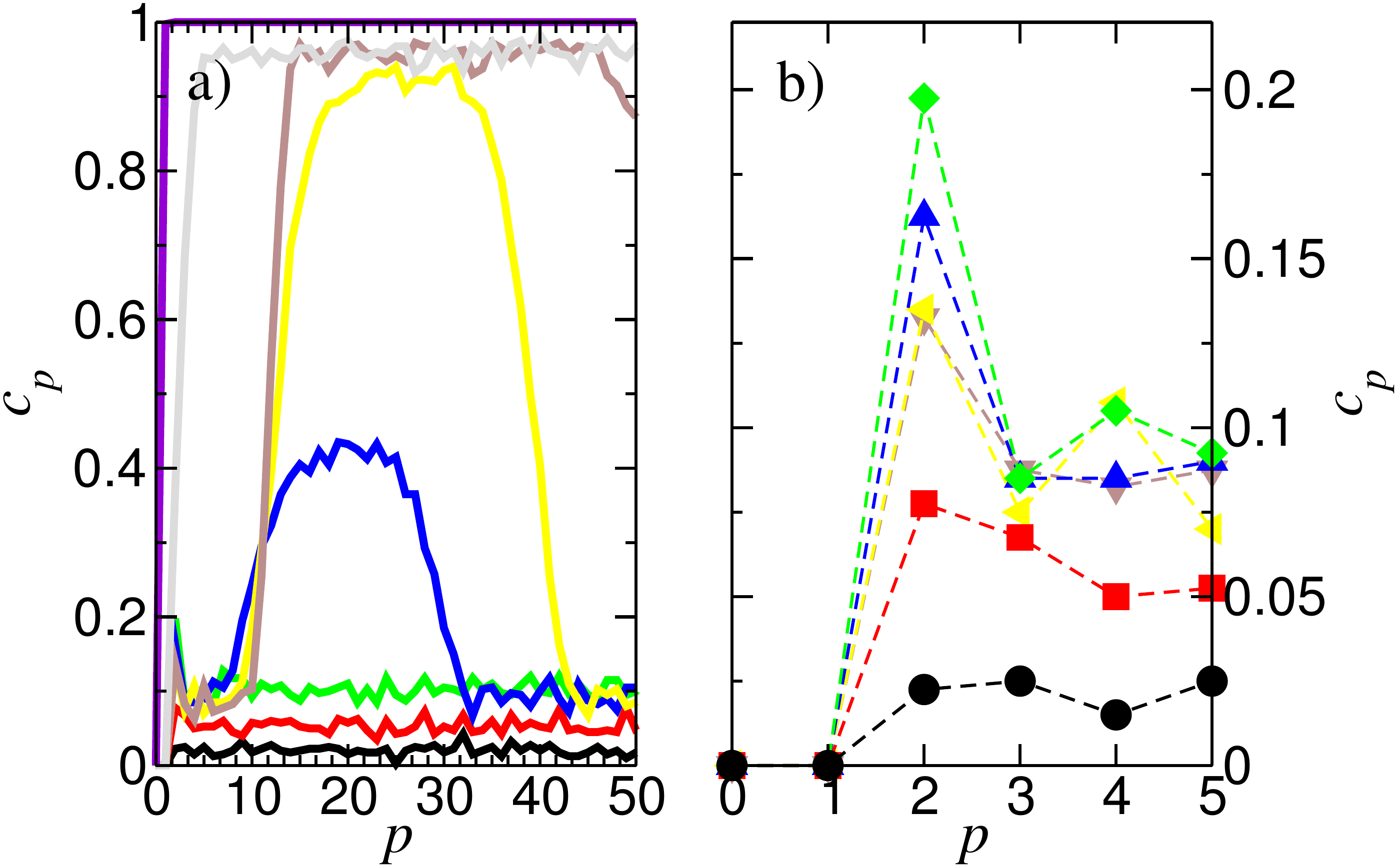}
\par\end{centering}

\caption{\label{fig:concentration_profile_36900}a)~and b)~Concentration
profiles of a FeCr surface at 300~K. Index $p=0$ indicates the top
surface layer. Various bulk concentrations \textbf{$c_{b}$ }are indicated:
0.02 (black); 0.05 (red); 0.10 (green); 0.15 (blue); 0.30 (yellow);
0.50 (brown); 0.90 (gray); 0.98 (violet); b)~Only the first six layers
are shown. Note the change in scale.}
\end{figure}

In figure~\ref{fig:concentration_profile_36900}, we plot the concentration
profile $c_{p}(p)$ of FeCr at 300~K ($\approx\nicefrac{1}{3}$~$T_{c}$)
and various bulk concentrations $c_{b}$ ranging from 0.02 to 0.98.
Special attention is given to the temperature range that is of industrial and technological
importance, \emph{i.e.} for $c_{b}$ below 0.3. The concentration profiles are highly non-linear
functions of the bulk concentration.

\begin{figure}
\begin{centering}
\includegraphics[width=1\columnwidth]{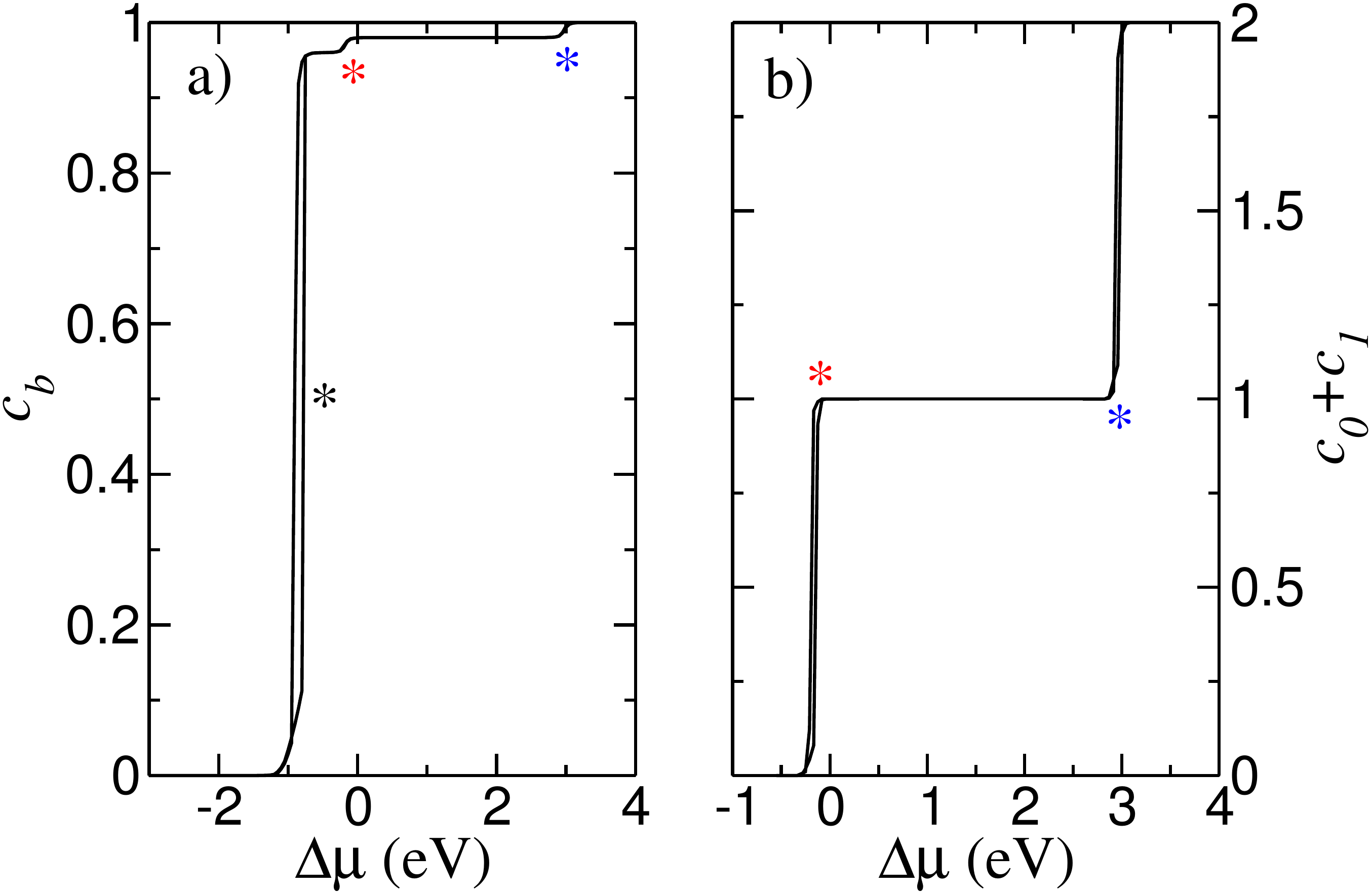}
\par\end{centering}

\caption{\label{fig:pseudoGP_c0}a)~Evolution of the bulk concentration with
the chemical potential $\Delta\mu$ at 300~K. b)~Evolution of the
sum of the concentrations of the top surface and subsurface layers, versus
the same chemical potential. In both figures, the black asterisk indicates the bulk
phase separation, the red and blue asterisks show the transitions in the subsurface and surface layers, respectively.}
\end{figure}

In order to get insight to these concentration profiles,
semi-grand canonical Monte Carlo simulations have been performed,
where the total number of sites and the difference in chemical potential
$\Delta\mu$ between pure bcc-iron and pure bcc-chromium are kept
fixed, while the bulk
concentration is free~\cite{ducastelle_order_1991}. The evolution of the surface
and bulk concentrations with respect to $\Delta\mu$ at 300~K are
plotted in Fig.~\ref{fig:pseudoGP_c0}. Three hysteresis loops are
found, which are indicated in the figure by asterisks. They bring out three phase transitions.
The first and stronger one, indicated by the black asterisk in Fig.~\ref{fig:pseudoGP_c0}.a.,
is an evidence of the well-known bulk phase-separation $\alpha-\alpha^{\prime}$
happening in bcc Fe-Cr alloys and discussed in the introduction. Note that the bulk solubility limit at low chromium concentration is slightly affected by the presence of the surface. It causes the large variations in the density profiles  in Fig.~\ref{fig:concentration_profile_36900} more than 10
layers away from the surface for $c_{b}\gtrapprox0.12$.
Snapshots of systems that undergo phase-separation at these concentrations are shown
in Fig.~\ref{fig:supercell}. 
%An important point is that this phase-separation
%does not occur at the expected bulk concentration, but at a higher
%value $c_{b}\approx0.12$. The presence of surfaces \emph{inhibits} the
%bulk phase transition. 
At higher difference in chemical potential, as indicated in Fig.~\ref{fig:pseudoGP_c0}.b.,
two less visible phase transitions occur. Each transition is localized in a single layer.
The first one corresponds to the subsurface layer transiting from pure Fe to pure
Cr (indicated by the red asterisk in Fig.~\ref{fig:pseudoGP_c0}.a.,
followed by an accompanying transition in the surface layer (blue
asterisk in Fig.~\ref{fig:pseudoGP_c0}a. and b.). The
change in concentration of the two first layers is abrupt
and discontinuous. It also gives insight to the emptiness in Cr
of these layers. First, the difference in surface chemical potential of the two elements, which is proportional
to the difference in surface energies of Fe and Cr, implies that Fe recover the layers where bonds are dangling. Indeed, surface energies of Fe are always lower than that of Cr for a given orientation. They range from 2.2 to 3.4 J/m$^2$ and 3.2 to 4.2 J/m$^2$ for iron and chromium, respectively \cite{blonski_structural_2007,ossowski_density_2008,levesque_thesis}.
Secondly, and in relation, the chemical potential of surface atoms is much modified by the surface,
which explains why the alloy do not phase separate at the same concentration than in the bulk.
One could see them as two new alloying elements only present in the surfaces.

\begin{figure}
\begin{centering}
\includegraphics[width=1\columnwidth]{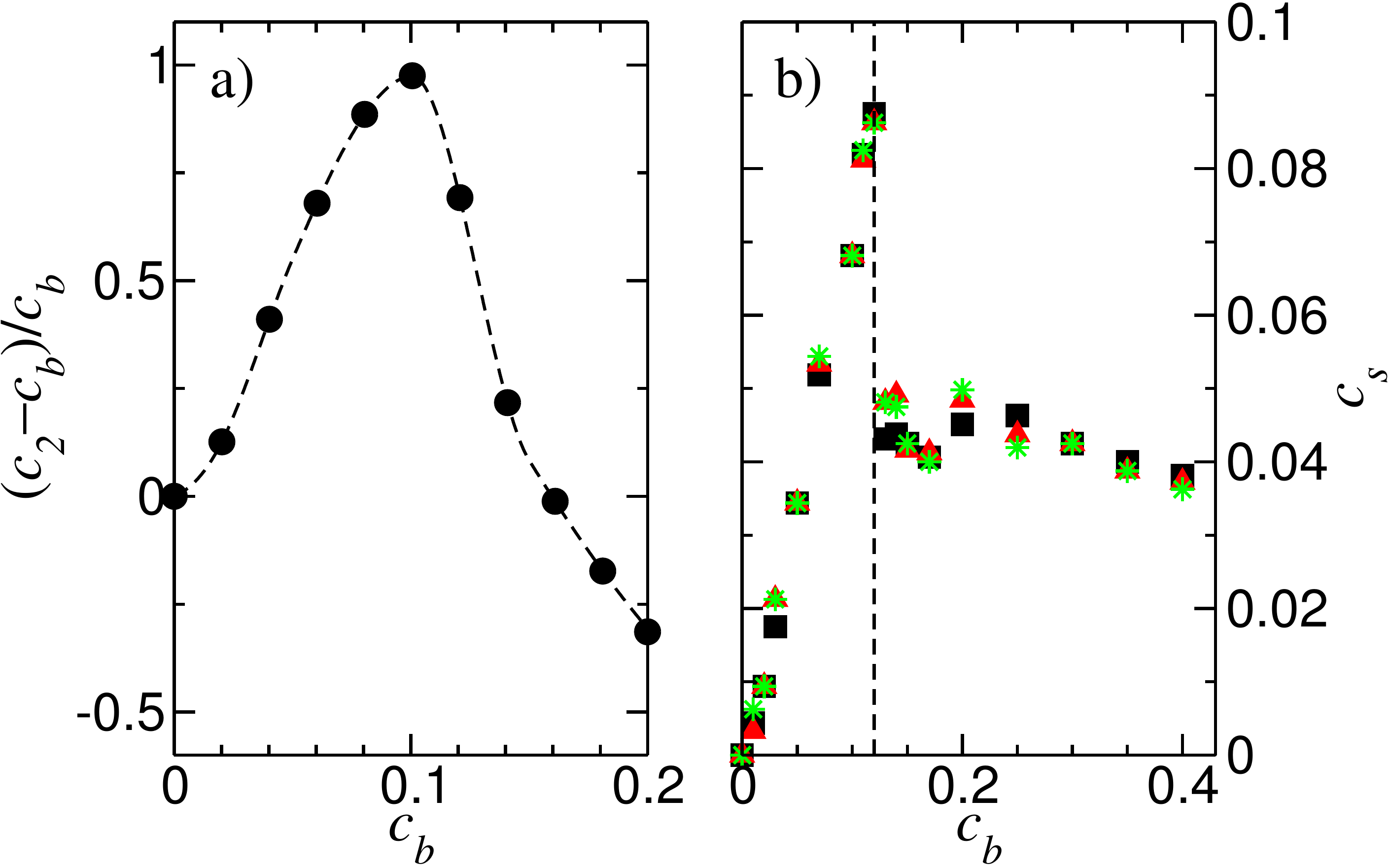}
\par\end{centering}

\caption{\label{fig:Profil_zoom_600K}a)~Relative evolution of the concentration
of the third layer, $c_{2}$, versus the bulk concentration, $c_{b}$,
at 300~K. b)~Average surface concentration, $\bar{c}_s=\left(c_{0}+c_{1}+c_{2}+c_{3}\right)/4$,
as a function of $c_{b}$ at 300~K (black squares), 450~K (red triangles)
and 600~K (green stars). A black shaded line indicates the optimal
concentration $c_{b}^{\text{opt}}\approx0.12$ at which ferritic FeCr steel
is the most corrosion resistant.}
\end{figure}

As shown in Fig.~\ref{fig:concentration_profile_36900}.b
and \ref{fig:Profil_zoom_600K}.a., the concentration of the
third layer, $c_{2}$, increases quickly with $c_{b}$, contrary to that of the first two layers discussed above: $c_2$ gets
from 0 to 0.2 when $c_{b}$ goes from 0 to 0.1, which represents a relative increase
in $c_{2}$ of about 100~\% at $c_{b}\approx0.1$. Importantly, $c_{2}$
is at this point greater than the bulk solubility limit, so that phase separation
would occur in the absence of the surface: its presence changes here the very nature of the Fe and Cr atoms, particularly their complex magnetic interactions as shown experimentally~\cite{unguris_magnetism_1992} and theoretically\cite{LevCrseg}, and consequently the perturbated alloy's thermodynamics. For higher bulk concentrations,
$c_{2}$ decreases sharply and becomes even depleted in Cr with respect
to the bulk at $c_{b}\gtrapprox0.13$.

Finally, and importantly, we plot in Fig.~\ref{fig:Profil_zoom_600K}.b the surface concentration, $\bar{c}_s$,
defined as the average concentration of all layers in direct contact
with the surface, \emph{i.e.}  $\bar{c}_s=\sum_{i=1}^{i_\text{max}}c_i/i_\text{max}$ with $i_\text{max}=4$ for orientation (100), as
a function of $c_{b}$ at 300~K, 450~K and 600~K.
Two regimes are clearly identified: (i) For bulk concentrations under
 0.12, the surface concentration increases with $c_{b}$, up to
a narrow maximum between 0.09 and 0.12. As soon as $c_{b}=0.07$,
surface composition exceeds the bulk solubility limit,
where phase-separation would occur in the bulk alloy. (ii) For larger
bulk compositions, there is a discontinuity in $c_{s}$, which is
reduced to a flat regime $c_{s}\approx0.05$, equivalent to that of
a $\alpha-\alpha^{\prime}$ phase-separated bulk. Indeed, 
$\alpha^{\prime}$ precipitation occurs as expected and discussed above to explain
the large variations in the density profiles of Fig.~\ref{fig:concentration_profile_36900},
and illustrated in Fig.~\ref{fig:supercell}.
\\Between 300~K and 600~K, the profiles shown in Fig.~\ref{fig:Profil_zoom_600K} are almost temperature-independent. 
It reflects a subtle compensation between the temperature dependence of the order energy, more specifically of the energy associated with magnetism,
 and the entropic effects. It was identified in the bulk as the cause of the
anomalously steep solubility limit of the Fe--Cr alloy at low temperatures, and identified by Williams as an effect of magnetism~\cite{williams_miscibility_1974,LevFeCr}.

An atomistic explanation of the thermodynamic origin of the
narrow optimum in corrosion resistance of stainless steels emerges from the above results.
The difference in surface chemical potential between Fe and Cr induces a strong Cr depletion in the first layers, where atoms have dangling bonds.
This result can be understood as a surface effect resulting from the surface energies of Fe being always lower than that of Cr.
 The sub-surface layers balance this local depletion by a strong enrichment in Cr, leading to three distinct regimes:
(i)~at low bulk concentrations, the ordering energy drives chromium into solution, far away from the surface: the resistance to corrosion is low and increases with Cr concentration.
(ii)~in a narrow range of bulk concentrations between 0.07 and 0.12, enough Cr is present to strongly enrich the surface in average but not to exceed the $\alpha$-$\alpha^\prime$ solubility limit\footnote{As discussed above, the $\alpha-\alpha^\prime$ phase separation occurs near surfaces at a solubility limit slightly different from that of the pure bulk alloy as studied in Ref.~\cite{LevFeCr}}: the Cr content in the surface is maximum.
%(ii)~a synergistic behavior happens in a narrow range of bulk concentrations between 0.07 and 0.12, where enough Cr is present
%to strongly enrich the surface in average but not to exceed the solubility limit in the bulk: the corrosion resistance is optimal.
(iii)~at higher bulk concentrations, the bulk solubility limit is exceeded and
most of available chromium bulk-precipitates in the $\alpha^{\prime}$ phase, depleting the surfaces: stainlessness is lost.

In light of the above results, in order to improve the stainlessness of
the Fe--Cr system and its derivative alloys, one may consider adding alloying elements that increase 
the solubility of Cr in Fe, without altering the anomalous thermodynamics of the Fe--Cr surfaces. The sharp decrease 
in the surface concentration would thus happen, nevertheless, but at higher bulk concentrations. Larger surface concentrations would be reached, inducing better protection against corrosion.
One could also increase the quantity of chromium atoms at the surfaces without increasing the content in the bulk. This could for instance be done by the localized addition of chromium-rich nanoscale precipitates or dispersoids. Such strategy is already under investigation as it is also a promising route to strengthen these materials~\cite{odette_annual_review_2008,brocq_2012}.

%\bibliographystyle{apsrev4-1}
%\bibliography{text}
%\end{document}

\end{document}